\documentclass{article}
\usepackage{epsfig}
\usepackage{amsfonts}
\usepackage{amsthm}
\usepackage{amssymb}
\usepackage{amscd}
\usepackage{graphicx}

\newcommand{\bastar}{\begin{eqnarray*}}
\newcommand{\eastar}{\end{eqnarray*}}
\newcommand{\be}{\begin{equation}}
\newcommand{\ee}{\end{equation}}
\newcommand{\bea}{\begin{eqnarray}}
\newcommand{\eea}{\end{eqnarray}}
\newcommand{\X}{{\vec X}}
\newcommand{\pro}{\partial}
\newcommand{\n}{\hat n}
\newcommand{\oneg}{\displaystyle\frac{1}{g}}

\newcommand{\D}{{\hat D}}

\newcommand{\A}{{\vec A}}
\newcommand{\valpha}{{\vec \alpha}}

\newcommand{\dfrac}{\displaystyle\frac}
\newcommand{\ba}{\begin{array}}
\newcommand{\ea}{\end{array}}

\newcommand{\nn}{\nonumber}

\begin{document}
\title{Isotropic and Gauge Invariant Monopole Condensation in SU(2) QCD}

\author{M. L. Walker\\ Institute of Quantum Science, 
College of Science and Technology,\\
Nihon University, Chiyoda, Tokyo 101-8308, Japan
\footnote{Email: m.walker@aip.org.au}}
\date{}

\maketitle
\abstract{I construct an isotropic, gauge invariant monopole condensate in $SU(2)$ QCD. It provides a suitable, gauge-invariant vacuum for the dual-Meissner effect to provide QCD colour 
confinement without violating the isotropy of real space.}


One of the most outstanding problems
in theoretical physics is the confinement problem in
quantum chromodynamics (QCD). It has
long been argued that monopole condensation can explain the
confinement of color through the dual Meissner effect \cite{N74,cho80a,Cho81}.
A magnetic condensate was first argued to be energetically favourable by
Savvidy \cite{S77}, but Nielsen and Olesen soon found an unstable mode
for such a background \cite{NO78}. This led some to consider variations such as the 
'spaghetti' vacuum which does indeed lower the energy to below that of the Savvidy
vacuum \cite{NO79} but whose construction requires the Nielsen and Olesen instability.
The plot thickens with other, more recent work \cite{H72,CP02,CmeP04,Cme04,sch,K04}
arguing that the Savvidy vacuum is stable and that the apparent instability of Nielsen and
Olesen is an artifact of the quadratic approximation \cite{KKP05}. 
Of course, not all such papers
have argued in these exact words. Cho \textit{et.al.} have used subtle causality and other
arguments \cite{CP02,CmeP04}, 
while Kondo \cite{K04} has argued that the condensate spontaneously generates 
a gluon mass to remove the instability. 

Identifying the form of the monopole condensate in QCD and proving its stability
has proved to be a very difficult task \cite{H72,CmeP04,K04,F83}. The first, and still most 
widely considered construction of the 
magnetic condensate by Savvidy \cite{S77} contains some real deficiences.
These include, lack of gauge invariance since the Savvidy vacuum selects the internal
$\hat{e}_3$ direction, real space anisotropy since it selects the $\hat{z}$ direction in
real space, and its inability to specify that the magnetic background is due to monopoles.

The first and last of these deficiences are solved by the 
Cho-Faddeev-Niemi-Shabanov (CFNS) decomposition, which I quickly review now \cite{cho80a,Cho81}.

There is a gauge-independent decomposition of the gauge potential into the restricted potential
$\hat A_\mu$ and the valence potential $\vec X_\mu$ \cite{cho80a,Cho81,FN99,S99}.
Let $\n$ be the unit
isovector which selects the color charge direction everywhere
in space-time, and let \cite{cho80a,Cho81}
\bea \label{eq:decompose}
& \vec{A}_\mu =A_\mu \n - \oneg \n\times\pro_\mu\n+\X_\mu\nonumber
         = \hat A_\mu + \X_\mu, \nn\\
& (A_\mu = \n\cdot \vec A_\mu,~ \n^2 =1,~ \hat{n}\cdot\vec{X}_\mu=0),
\eea
where $A_\mu$ is the ``electric'' potential.
Notice that the restricted potential $\hat A_\mu$ is precisely
the connection which leaves $\n$ invariant under parallel transport,
\bea
\D_\mu \n = \pro_\mu \n + g {\hat A}_\mu \times \n = 0.
\eea
This paper is only concerned with $SU(2)$ but
the generalisation to arbitrary numbers of colours has been done \cite{C80,FN99c,LZZ00}.
Under the infinitesimal gauge transformation
\bea
\delta \n = - \vec \alpha \times \n  \,,\,\,\,\,
\delta \A_\mu = \oneg  D_\mu \vec \alpha,
\eea
one has
\bea \label{eq:decomptrans}
&&\delta A_\mu = \oneg \n \cdot \pro_\mu \valpha,\,\,\,\
\delta \hat A_\mu = \oneg \D_\mu \valpha  ,  \nn \\
&&\hspace{1.2cm}\delta \X_\mu = - \valpha \times \X_\mu  .
\eea
This tells that $\hat A_\mu$ by itself describes an $SU(2)$
connection which enjoys the full $SU(2)$ gauge degrees of
freedom. Furthermore the valence potential $\vec X_\mu$ forms a
gauge covariant vector field under the gauge transformation.

The important thing for this article is that the decomposition is
gauge independent. Once the gauge covariant topological field
$\hat n$ is chosen, the decomposition follows automatically,
regardless of the choice of gauge \cite{cho80a,Cho81}.
Furthermore, $\hat{A}_\mu$ retains all the essential
topological characteristics of the original non-Abelian potential.
Clearly $\hat{n}$ defines $\pi_2(S^2)$
which describes the non-Abelian monopoles \cite{WY75,cho80}, and
characterizes
the Hopf invariant $\pi_3(S^2)\simeq\pi_3(S^3)$ which describes
the topologically distinct vacua \cite{BPST75,cho79}.

$\hat{A}_\mu$ has a dual
structure,
\begin{eqnarray}
& \hat{F}_{\mu\nu} = \partial_\mu \hat A_\nu-\partial_\nu \hat A_\mu
+ g \hat A_\mu \times \hat A_\nu = (F_{\mu\nu}+ H_{\mu\nu})\hat{n}\mbox{,}
\nonumber \\
& F_{\mu\nu} = \partial_\mu A_{\nu}-\partial_{\nu}A_\mu \mbox{,}
\nonumber \\
& H_{\mu\nu} = -\dfrac{1}{g} \hat{n}\cdot(\partial_\mu
\hat{n}\times\partial_\nu\hat{n}),
\end{eqnarray}
One can identify the non-Abelian monopole potential by
\bea \label{eq:vecC}
\vec B_\mu= -\dfrac{1}{g}\hat n \times \partial_\mu\hat n ,
\eea
in terms of which the magnetic field is expressed by
\bea
\vec H_{\mu\nu}=\partial_\mu \vec B_\nu-\partial_\nu \vec B_\mu+ g
\vec B_\mu \times \vec B_\nu = H_{\mu\nu}\hat n.
\eea
This provides the gauge independent separation of the monopole
field $H_{\mu\nu}$ from the color electromagnetic field $ F_{\mu\nu}$.

This just leaves the issue of isotropy
invariance. The QCD vacuum should be isotropy invariant and 
this could be seen as another argument for the spaghetti vacuum 
with its magnetic flux lines running every which way. The purpose of this note is 
to construct a stable isotropic monopole condensate.

The field at any given point
resulting from the CFNS construction can always be Lorentz transformed
to a pure magnetic field \cite{cho80a} because of its monopole origin, just as that
due to electric charges can always be Lorentz transformed to a purely electrical one.
I therefore work in such a reference frame from the outset and take all $\vec H_{0\nu}=0$.
Considering a magnetic background of the form
\be \label{eq:hedgehog}
\vec{H}_{ij} = \epsilon_{ijk} \hat{n}^k H,
\ee
where
\be
H = \hat{n}^p \epsilon_{lmp} \partial_l \hat{n} \times \partial_m \hat{n}
\ee
is the magnitude of the field strength.
This is essentially a hedgehog monopole solution because the direction in real space
has been linked with the direction in the internal space. The important thing for the
purpose of this paper however is that if the monopole field is non-zero then its 
direction in real space must vary continuously. 
In this way it achieves isotropy in the same way it
achieves gauge invariance. Note that while both these degrees of freedom are no longer
fixed as in the conventional Savvidy vacuum, they are now linked so there is still some
loss of freedom. This is in fact desirable since it guarantees that travelling around a
closed loop in real space leaves the field in the same internal direction, as required
for a well-defined field.
Finally, we know that such a background field configuration mathematically exists 
because it is locally the Savvidy vacuum everywhere in spacetime.

The construction may be criticised for lacking complete Lorentz invariance. 
It is tempting to speculate that the physical vacuum is a superposition of such vacuum
states, each Lorentz transformed to its own preferred frame. However,
while the electric and magnetic components of an electromagnetic field can be found by
observing the trajectories of three or more charged particles, such a measurement is not
feasible in QCD whose colour charges are both confined and bound, so it is not clear that
one could actually determine a preferred reference frame. It is also worth remembering that 
the main
interest of these studies is confinement via the dual superconductor mechanism, and this
depends only on the magnitude of the field strength, which is Lorentz invariant.

The greatest deficiency in the author's view is that it is not easily adapted to 
three-colour QCD. The importance of this is obvious. An examination of $SU(3)$ hedgehog
solutions is the obvious place to start looking.

I have argued for the existence of a stable, isotropic Savvidy-like vacuum capable of
causing confinement via the dual superconductor mechanism in two-colour QCD. Attempts to
generalise this to three-colour QCD are ongoing.


\begin{thebibliography}{99}
\bibitem{N74}Y. Nambu, 
\textit{Strings, Monopoles and Gauge Fields}, Phys.\ Rev.\ \textbf{D10}, 4262 (1974);
S. Mandelstam, 
\textit{Vortices And Quark Confinement In Nonabelian Gauge Theories},
Phys.\ Rep.\ \textbf{23C}, 245 (1976);
A. Polyakov, \textit{Quark Confinement And Topology Of Gauge Groups},
Nucl.\ Phys.\ \textbf{B120}, 429 (1977);
G. 't Hooft, \textit{Topology Of The Gauge Condition And New Confinement Phases In Nonabelian Gauge Theories}, Nucl.\ Phys.\ \textbf{B190}, 455 (1981).
\bibitem{cho80a}Y.M. Cho, \textit{A Restricted Gauge Theory},
Phys.\ Rev.\ \textbf{D21}, 1080 (1980); J.\ Korean Phys.\ Soc.\ \textbf{17}, 266 (1984); 
Phys.\ Rev.\ \textbf{D62}, 074009 (2000).
\bibitem{Cho81}Y.M. Cho, Phys.\ Rev.\ Lett.\ \textbf{46}, 302 (1981); 
\textit{Extended Gauge Theory And Its Mass Spectrum}, Phys.\ Rev.\ \textbf{D23}, 2415 (1981); 
W.S. Bae, Y.M. Cho, and S.W. Kimm, \textit{QCD versus Skyrme-Faddeev theory}, Phys.\ Rev.\ \textbf{D65}, 025005 (2002).
\bibitem{S77} G.K. Savvidy, \textit{Infrared Instability Of The Vacuum State Of Gauge Theories And Asymptotic Freedom}, Phys.\ Lett.\ \textbf{B71}, 133 (1977).
\bibitem{NO78} N. Nielsen and P. Olesen, \textit{An Unstable Yang-Mills Field Mode}, Nucl.\ Phys.\ \textbf{B144}, 376 (1978);
C. Rajiadakos, \textit{A Stable Symmetrized Savvidy Vacuum}, 
Phys.\ Lett.\ \textbf{B100}, 471 (1981).
\bibitem{NO79} N. Nielsen and P. Olesen, \textit{A Quantum Liquid Model for the QCD Vacuum}, Nucl.\ Phys.\ \textbf{B160}, 380 (1979);
\bibitem{H72} J. Honerkamp, Nucl. Phys. {\bf B48}, 269 (1972).
\bibitem{CP02}Y.M. Cho and D.G. Pak, \textit{Monopole condensation in SU(2) QCD},
Phys.\ Rev.\ \textbf{D65},074027 (2002).
\bibitem{CmeP04}
Y.M. Cho, M.L. Walker, and D.G. Pak, {\it Monopole condensation and
  confinement of color in su(2) qcd},  JHEP {\bf 05}, 73 (2004), hep-th/0209208.
\bibitem{Cme04} Y.M. Cho and M.L. Walker, 
\textit{Stability of Monopole Condensation in SU(2) QCD}, hep-th/0206127.
\bibitem{sch} V. Schanbacher, 
\textit{Gluon Propagator And Effective Lagrangian In QCD}, Phys.\ Rev.\ \textbf{D26},489 (1982);
L. Freyhult, \textit{Field Decomposition and the Ground State Structure of SU(2) Yang-Mills Theory}, Int.\ J.\ Mod.\ Phys.\ \textbf{A17}, 3681 (2002), hep-th/0106239.
\bibitem{K04}
K.-I. Kondo, {\it Magnetic condensation, abelian dominance and instability of
  savvidy vacuum},  Phys.\ Lett.\ \textbf{B600}, 287 (2004), hep-th/0404252.
\bibitem{KKP05}
D.~Kay, A.~Kumar, and R.~Parthasarathy, {\it Savvidy vacuum in su(2) yang-mills
  theory}, Mod.\ Phys.\ Lett.\ {\bf A20}, 1655 (2005).
\bibitem{F83}C. Flory, \textit{A Selfdual Gauge Field, Its Quantum Fluctuations, And Interacting Fermions}, Phys.\ Rev.\ \textbf{D28}, 1425 (1983);
\bibitem{FN99} L. Faddeev and A. Niemi, 
\textit{Partially dual variables in SU(2) Yang-Mills theory}, 
Phys.\ Rev.\ Lett.\ \textbf{82}, 1624 (1999); 
\textit{Partial duality in SU(N) Yang-Mills theory}, Phys.\ Lett.\ \textbf{B449}, 214 (1999).
\bibitem{S99}S. Shabanov, \textit{An effective action for monopoles and knot solitons in Yang-Mills  theory}, Phys.\ Lett.\ \textbf{B458}, 322 (1999);
\textit{Yang-Mills theory as an Abelian theory without gauge fixing},
Phys.\ Lett.\ \textbf{B463}, 263 (1999); H. Gies, 
\textit{Wilsonian effective action for SU(2) Yang-Mills theory with  Cho-Faddeev-Niemi-Shabanov decomposition}, Phys.\ Rev.\ \textbf{D63}, 125023 (2001);
R. Zucchini, 
\textit{Global Aspects of Abelian and Center Projections in SU(2) Gauge Theory}, hep-th/0306287.
\bibitem{C80}
Y.~M. Cho, {\it Colored monopoles},  Phys.\ Rev.\ Lett.\ {\bf 44}, 1115 (1980).

\bibitem{FN99c}
L.~D. Faddeev and A.~J. Niemi, {\it Partial duality in su(n) yang-mills
  theory}, Phys.\ Lett.\ {\bf 449}, 214 (1999), hep-th/9812090.

\bibitem{LZZ00}
S.~Li, Y.~Zhang, and Z.-y. Zhu, {\it Decomposition of su(n) connection and
  effective theory of su(n) qcd}, Phys.\ Lett.\ {\bf 487}, 201 (2000), hep-th/9911132.
\bibitem{WY75}T.T. Wu and C.N. Yang, \textit{Concept Of Nonintegrable Phase Factors And Global Formulation Of Gauge  Fields}, Phys.\ Rev.\ \textbf{D12}, 3845 (1975).
\bibitem{cho80} Y.M. Cho, \textit{Colored Monopoles}, Phys.\ Rev.\ Lett.\ 
\textbf{44}, 1115 (1980); 
\textit{Internal Structure Of The Monopoles}, Phys.\ Lett.\ \textbf{B115}, 125 (1982).
\bibitem{BPST75}A. Belavin, A. Polyakov, A. Schwartz, and Y. Tyupkin,
\textit{Pseudoparticle Solutions Of The Yang-Mills Equations},
Phys.\ Lett.\ \textbf{B59}, 85 (1975); 
G. 't Hooft, \textit{Symmetry Breaking Through Bell-Jackiw Anomalies}, 
Phys.\ Rev.\ Lett.\ \textbf{37}, 8 (1976).
\bibitem{cho79}Y.M. Cho, 
\textit{Vacuum Tunneling In Spontaneously Broken Gauge Theory}, 
Phys.\ Lett.\ \textbf{B81}, 25 (1979).
\bibitem{F80} H.~Flyvbjerg,
\textit{Improved qcd vacuum for gauge groups su(3) and su(4)},
Nucl. Phys. {\bf B176}, 379 (1980).
\bibitem{me07}
M.L. Walker,
{\it Stability of the magnetic monopole condensate in three- and
  four-colour qcd}, JHEP {\bf 01}, 056 (2007).

\end{thebibliography}
\end{document}